\documentclass{appolb}
\usepackage{graphicx}
\usepackage{amsmath}

\def\be{\begin{equation}}
\def\ee{\end{equation}}
\newcommand{\beq}{\begin{equation}}
\newcommand{\eeq}{\end{equation}}
\newcommand{\bea}{\begin{align}}
\newcommand{\eea}{\end{align}}
\newcommand{\bsp}{\begin{equation}\begin{split}}
\newcommand{\esp}{\end{split}\end{equation}}
\def\beq{\begin{equation}}
\def\eeq{\end{equation}}
\def\bsp#1\esp{\begin{split}#1\end{split}}

\newcommand{\cA}{\begin{cal}A\end{cal}}

\newcommand{\cF}{\begin{cal}F\end{cal}}

\newcommand{\cN}{\begin{cal}N\end{cal}}

\newcommand{\cR}{\begin{cal}R\end{cal}}

\newcommand{\zbar}{{\bar z}}
\newcommand{\zb}{{\bar z}}
\newcommand{\bx}{\textbf{x}}
\newcommand{\bp}{\textbf{p}}
\newcommand{\bq}{\textbf{q}}

\begin{document}
\title{Amplitudes in the Multi-Regge Limit of $\mathcal{N}=4$ SYM%
\thanks{Presented at Low X and Diffraction 2018.}%
}
\author{Vittorio Del Duca$^{a}$,
Stefan Druc$^b$,
James Drummond$^b$,\\
Claude Duhr$^{c,d}$,
Falko Dulat$^e$,
Robin Marzucca$^d$,\\
Georgios Papathanasiou$^f$,
Bram Verbeek\footnote{Speaker.}$^{,d}$
\address{${}^a$ Institute for Theoretical Physics, ETH Z\"urich, 8093 Z\"urich, Switzerland.\\
${}^b$ School of Physics \& Astronomy, University of Southampton, \\
Highfield, Southampton, SO17 1BJ, United Kingdom.\\
${}^c$ Theoretical Physics Department, CERN, CH-1211 Geneva 23, Switzerland.\\
${}^d$ Center for Cosmology, Particle Physics and Phenomenology (CP3),\\
UCLouvain, Chemin du Cyclotron 2, 1348 Louvain-La-Neuve, Belgium.\\
${}^e$ SLAC National Accelerator Laboratory, Stanford University, \\
Stanford, CA 94309, USA.\\
${}^f$ DESY Theory Group, DESY Hamburg,\\
 Notkestra{\ss}e 85, D-22607 Hamburg, Germany.}
 }

\maketitle

\preprint{CERN-TH-2018-252,CP3-18-64,DESY 18-190}
\prepNo

\begin{abstract}
A novel way of computing high-order amplitudes in the multi-Regge limit of planar maximally supersymmetric Yang-Mills theory is presented. In this framework, we are able to obtain high-loop and high-leg results by an easy operation on known amplitudes with fewer loops and lower multiplicity. This mechanism will be reviewed, along with an ensuing factorisation which allows us to determine leading logarithmic MHV results for any number of legs at a fixed loop order.
\end{abstract}

\section{Introduction}

Maximally supersymmetric Yang-Mills theory in four dimensions ($ \cN = 4$ SYM) is one of the most studied models in modern physics. In recent years, many properties have been uncovered which led to insights on the perturbative regime of the theory.  For one, beyond the ordinary conformal symmetry which $\mathcal{N}=4$ SYM exhibits, a dual conformal symmetry \cite{Drummond:2006rz,Bern:2006ew,Bern:2007ct,Alday:2007hr,Drummond:2007aua} was discovered in the planar limit. This symmetry fixes the 4 and 5-particle case completely to the so-called Bern-Dixon-Smirnov (BDS) ansatz \cite{Bern:2005iz}. Beyond 5 particles there is an additional non-trivial dual conformally invariant contribution \cite{Bern:2008ap}. Dual conformal symmetry also spawned a new approach to kinematics in terms of momentum twistors \cite{Hodges:2009hk}. This reformulation has led to a deeper understanding of the function space of amplitudes in planar $\mathcal{N} = 4$ SYM for the MHV and NMHV case \cite{ArkaniHamed:2012nw} and their singularity structure in general \cite{Golden:2013xva}.\\
Due to these developments, many impressive results in planar $\mathcal{N}=4$ SYM at high loop orders with many particles were obtained. The 6-particle amplitude is known at 5 loops \cite{Caron-Huot:2016owq}. At 7 particles the MHV amplitude is known analytically at 2 loops \cite{Golden:2014xqf} and at symbol level up to 4 loops in the MHV case and at 3 loops in the NMHV case  \cite{Dixon:2016nkn}. However, the techniques used to obtain these results are currently limited to 7 or fewer external particles, and further study of the high-particle case is required to move beyond the state of the art. One way to achieve this goal is by studying a special kinematic limit. In what follows, we will study amplitudes in planar $ \cN=4$ SYM in the multi-Regge limit.


\section{Amplitudes in the Multi-Regge Limit}

Let us introduce the multi-Regge limit. Consider $1 \, + \, 2 \rightarrow 3 \, + \, \dots + N$ scattering with all particles outgoing and introduce lightcone coordinates:
\begin{equation}
p^\pm \equiv p^0 \pm p^z, \qquad \bp_{k} \equiv p_{k\perp} = p_k^x + i p_k^y \,.
\end{equation}
If we choose the reference frame in which the momenta of the initial state gluons lie on the $z$-axis with $p_2^0 = p_2^z$, implying $p_1^+ = p_2^- = \bp_1 = \bp_2 = 0$, the multi-Regge limit corresponds to the limit where
\begin{equation}
p_3^+ \gg p_4^+ \gg \dots \gg p_{N-1}^+ \gg p_N^+, \qquad |\bp_3| \simeq \dots \simeq |\bp_N| \,.
\end{equation}
We introduce dual coordinates $x_i$ as $x_i - x_{i-1} = p_i$.
Amplitudes in planar $\cN = 4$ SYM obey dual conformal invariance, which implies that the kinematical dependence can be expressed in terms of conformal cross-ratios of the dual coordinates. Of these, only $3N-15$ are algebraically independent in four dimensions. In the multi-Regge limit, where amplitudes only depend non-trivially on the transverse momenta $\bp_i$, this number of independent complex cross ratios reduces to $N-5$. The set of independent transverse cross ratios $z_i$ that we will use is defined as
\begin{equation}\label{eq:z_i_def}
z_i\equiv  \frac{({\textbf x}_1 -{\textbf x}_{i+3})\,({\textbf x}_{i+2} -{\textbf x}_{i+1})}{({\textbf x}_1 -{\textbf x}_{i+1})\,({\textbf x}_{i+2} -{\textbf x}_{i+3})}  = -\frac{{\textbf q}_{i+1}\,{\textbf k}_{i}}{{\textbf q}_{i-1}\,{\textbf k}_{i+1}}\,,
\end{equation}
with transverse dual coordinates
${\mathbf q}_i = {\mathbf x}_{i+2} - {\mathbf x}_1$ and ${\mathbf k}_i = {\mathbf x}_{i+2} - {\mathbf x}_{i+1}$.

With these kinematical considerations in mind, we turn to the form of amplitudes in multi-Regge kinematics (MRK) in planar $\cN =4 $ SYM. As mentioned before, dual conformal symmetry fixes the 4 and 5 point amplitude to all orders but beyond 5 points the amplitude diverges from the BDS ansatz and we get an additional dual conformally invariant contribution.
In the multi-Regge limit helicity is conserved by the gluons going very forward, and thus the helicity configuration is determined by the produced gluons exclusively. Denoting these helicities by $h_1,\dots,h_{N-4}$, we define the ratio
\begin{equation}
e^{i\Phi_{h_1,\dots,h_{N-4}}}\cR_{h_1,\dots,h_{N-4}} \equiv \frac{\cA_N(-,+,h_1,\dots,h_{N-4},+,-)}{\cA^{\text{BDS}}_N(-,+,h_1,\dots,h_{N-4},+,-)}\bigg|_{\text{MRK}}
\end{equation}
where $e^{i\Phi_{h_1,\dots,h_{N-4}}}$ is a phase factor such that in the Euclidian region we have $\cR_{h_1,\dots,h_{N-4}} =1$. When performing an analytic continuation to another Mandelstam region, $\cR_{h_1,\dots,h_{N-4}}$ picks up contributions called Regge cuts. In the Mandelstam region where the energy components of all produced particles are analytically continued, it can be written as a dispersion integral of a product of building blocks. In the multi-Regge limit large logarithms appear, resummed to leading logarithmic accuracy (LLA) in these large logarithms, $\cR_{h_1,\dots,h_{N-4}}$ is given by \cite{Bartels:2011ge,DelDuca:2016lad}
\begin{equation}
\begin{split}\label{eq:MRK_conjecture}
\cR_{h_1,\dots,h_{N-4}}= &1  + a\,i\pi\, \left[ \prod_{k=1}^{N-5}\sum_{n_k=-\infty}^{+\infty}\left(\frac{z_k}{\zbar_k}\right)^{\frac{n_k}{2}}\int_{-\infty}^{+\infty}\frac{d\nu_k}{2\pi}|z_k|^{2i\nu_k} \right] \\
&\times\left[\prod_{k=1}^{N-5}e^{a \, \log (\tau_k) E_k}\right]\, \chi^{h_1}_{1} \left[\prod_{k=2}^{N-5}C^{h_{k}}_{k-1,k} \right] \,\chi^{-h_{N-4}}_{N-5}.
\end{split}
\end{equation}
Here,
$
\tau_k \equiv \delta_k \sqrt{\frac{|\bq_{k-1}|^2|\bq_{k+1}|^2|\bp_{k+3}|^2}{|\bq_k|^4|\bp_{k+4}|^2}}$ where in the multi-Regge limit we have $\delta_k = p^+_{k+4}/p^+_{k+3} \xrightarrow[\text{MRK}]{ } 0,
$
 and thus $\log \tau_k$ denotes the large logarithm. Furthermore, $a$ is the 't Hooft coupling and $\chi^{h_i}_k \equiv \chi^{h_i}(\nu_k,n_k)$,\linebreak  $ C^{h_{k}}_{k-1,k} \equiv C^{h_{k}}(\nu_{k-1},n_{k-1},\nu_k,n_k)$ and $ E_k \equiv E(\nu_k,n_k)$ are the leading order parts of building blocks called the impact factor, the central emission block and the BFKL eigenvalue respectively. The product of these building blocks is put through an integral transform called the Fourier-Mellin transform which is given by
 \begin{equation}
 \cF[f](z) = \sum_{n=-\infty}^{+\infty}\left(\frac{z}{\zbar}\right)^{\frac{n}{2}}\int_{-\infty}^{+\infty}\frac{d\nu}{2\pi}\,|z|^{2i\nu} f(\nu,n).
 \end{equation}
We are interested in the perturbative expansion of eq. \eqref{eq:MRK_conjecture}. We may write it at $\ell$ loops as
\begin{equation}
\begin{split}
\cR_{h_1,\dots,h_{N-4}}^{(\ell)}=2 \pi i a^\ell \sum_{\sum i_k = \ell -1} \left( \prod_{k=1}^{N-5} \frac{\log^{i_k} \tau_k}{i_k!} \right) g_{h_1,\dots,h_{N-4}}^{(i_1,\dots,i_{N-5})}(\{z_i\}).
\end{split}
\end{equation}
The objects $g_{h_1,\dots,h_{N-4}}^{(i_1,\dots,i_{N-5})}$ will henceforth be referred to as perturbative coefficients and are given by an $(N-5)$-fold Fourier-Mellin transform
\begin{equation}
  g_{h_1,\dots,h_{N-4}}^{(i_1,\dots,i_{N-5})} = \cF_{N-5}\left[ \chi^{h_1}_{1} \left(\prod_{k=2}^{N-5}C^{h_{k}}_{k-1,k} \right) \,\chi^{-h_{N-4}}_{N-5} \left(\prod_{l=1}^{N-5}E_l^{i_l} \right) \right].
\end{equation}
  In \cite{Dixon:2012yy,DelDuca:2016lad} it was shown that in momentum space these objects are made up of functions called multiple polylogarithms
\beq
G(a_1,\ldots,a_n; z) = \int_0^z\frac{dt}{t-a_1}G(a_2,\ldots,a_n;t)\,, \quad G(;z)=1,
\eeq
in combinations such that their branch cuts cancel. In what follows, we will study the Fourier-Mellin transform in some more detail to use its mathematical properties to facilitate the computation of the perturbative coefficients.

\section{Perturbative Coefficients through Convolutions}

The objects we wish to compute are Fourier-Mellin transforms of products of building blocks.
The Fourier-Mellin transform maps products into convolutions, so that for $\cF[F]=f$ and $\cF[G]=g$ we have $\cF[F\cdot G] = \cF[F]\ast\cF[G] = f\ast g$
where the convolution is given by
\begin{align}\label{eq:conv_def}
(f\ast g)(z) = \frac{1}{\pi}\int \frac{d^2w}{|w|^2}\,f(w)\,\,g\left(\frac{z}{w}\right)\,.
\end{align}
Thus, rather than recomputing the Fourier-Mellin integral for every perturbative coefficient, we might hope to use this convolution product to compute the perturbative orders recursively. Take, for instance a three-loop coefficient at seven points
  $g^{(1,1)}_{h_1,h_2,h_3}=\cF_{2}\left[ \chi^{h_1}_{1} C^{h_{2}}_{1,2} \chi^{-h_{3}}_{2} E_1 E_2 \right] = \cF\left[E_1 \right]*g^{(0,1)}_{h_1,h_2,h_3}$,
we see that it can be computed by convolution of a two-loop seven-point coefficient. Since the perturbative coefficients are single-valued \cite{Dixon:2012yy}, the evaluation of the convolution integrals can be simplified to a residue computation, as was shown in \cite{Schnetz:2013hqa}. Let $f(z)$ be a single-valued combination of polylogarithms over rational functions with singularities at $z=a_i$ and $z = \infty$.
Define the {holomorphic residue} of $f$ at $z = a$ as the coefficient of the simple holomorphic pole with no logarithmic singularities.
The integral of $f$ over the whole complex plane, if it exists, is given by the sum of the holomorphic residues of its single-valued antiholomorphic primitive $F$, i.e. if $\bar{\partial} F = f$, then
\begin{equation}
\int \frac{d^2z}{\pi}\,f(z) = \textrm{Res}_{z=\infty}F(z) - \sum_i\text{Res}_{z=a_i}F(z)\,.
\end{equation}
Thus, single-valuedness allows us to compute these convolution integrals easily. Further noting that the Fourier-Mellin transform of the BFKL eigenvalue is a rational function
$
  \cF[E_i] = -(z_i+\zb_i)/(2|1-z_i|^2),
$
we see that it is easy to go up in loop order recursively via
\begin{equation}
  g^{(i_1,\dots,i_k+1,\dots,i_{N-5})}_{h_1,\dots,h_{N-4}} = \cF\left[E_{i_k} \right]*g^{(i_1,\dots,i_k,\dots,i_{N-5})}_{h_1,\dots,h_{N-4}},
\end{equation}
as the perturbative coefficients are made up of multiple polylogarithms, and their integration over rational kernels are easy to compute. We can also use this method to change helicities. For example, take an NMHV three-loop coefficient at seven points
\begin{equation}
  g^{(1,1)}_{-,+,+}=\cF_{2}\left[ \chi^{-}_{1} C^{+}_{1,2} \chi^{-}_{2} E_1 E_2 \right] = \cF\left[ \frac{\chi^{-}_{1}}{\chi^{+}_{1}} \right]*g^{(1,1)}_{+,+,+},
\end{equation}
the extracted term is given by
$
  \cF\left[\chi^{-}_{i}/\chi^{+}_{i} \right] = -(z_i)/(1-z_i)^2,
$
and thus we can obtain perturbative coefficients beyond MHV by convoluting their MHV counterparts with a rational function. In fact, one can show that convolutions of this sort allow you to obtain all different helicity configurations.

Using this method, we may start from the known 2-loop MHV amplitude for any number of points \cite{Bartels:2011ge} and move our way through all helicities and up in loop number recursively. However, there is even more to be learned from convolutions. In fact, this method led to the discovery of a factorisation of the LLA perturbative coefficients \cite{DelDuca:2016lad}, which for the MHV case states that if we express the coefficients in terms of the transverse dual coordinates $\{ \bx_i \}$
\beq\bsp
g_{+,\ldots,+}^{(0,\ldots,0,i_{a_1},0,\ldots,0,i_{a_2},0,\ldots,0,i_{a_k},0,\ldots,0)}(\mathbf{x}_1,\ldots,\mathbf{x}_{N-5})\\=	g_{+,\ldots,+}^{(i_{a_1},i_{a_2},\ldots,i_{a_k})}(\mathbf{x}_{a_1},\ldots,\mathbf{x}_{a_k}).
\esp\eeq
Since at leading logarithmic accuracy $\sum_j i_{a_j} = \ell -1$, the factorisation allows us to determine MHV LLA amplitudes with any number of external legs, from the set of amplitudes with up to $\ell +4$ external legs.

\section{Conclusion}

We have presented a framework to compute scattering amplitudes in the multi-Regge limit of planar $\cN=4$ SYM efficiently. Using these methods, higher-loop contributions can be obtained by recursive operation on lower-loop results and helicity configurations beyond MHV can be obtained from MHV results. We have also presented a factorisation which allows us to determine scattering amplitudes for any number of particles by computing a finite number of perturbative coefficients.

These methods were first presented in \cite{DelDuca:2016lad} and allowed for the computation of the LLA MHV amplitude for any number of particles to 5 loops, the LLA amplitude for 8 or less particles for any helicity configuration up to 4 loops. In \cite{DelDuca:2018hrv} a first extension beyond leading logarithmic accuracy was considered which led to the 7 point amplitude at next-to-leading logarithmic accuracy (NLLA) through 5 loops for the MHV case, and through 3 and 4 loops for the two independent NMHV helicity configurations, respectively. A study beyond leading logarithmic accuracy for an arbitrary number of particles with an extension of the factorisation beyond LLA was presented in \cite{FutureCite}, which was applied to the 8 particle case in \cite{FutureCiteUs} to give the 8 point NLLA amplitude for any helicity configuration at 3 loops. In addition, the same formalism was applied beyond $\cN = 4$ SYM to the computation of the BFKL ladder at NLLA in \cite{DelDuca:2017peo}.


\section{Acknowledgements}
This work is supported by the European Research Council (ERC) through the grants 637019 (MathAm) and 648630 (IQFT), and by the U.S. Department of Energy (DOE) under contract DE- AC02-76SF00515.

\bibliographystyle{JHEP}
\bibliography{refs}

\end{document}